%% file: 0_main.tex
\documentclass[lettersize,journal]{IEEEtran}

\IEEEoverridecommandlockouts
\usepackage{cite}
\usepackage{amsmath,amssymb,amsfonts}
\usepackage{url}
\usepackage{algorithmic}
\usepackage{algorithm} 
\usepackage[french,boxed,vlined,linesnumbered,inoutnumbered,rightnl,algo2e]{algorithm2e}
\usepackage{graphicx}
\usepackage{textcomp}
\usepackage{xcolor}
\usepackage{comment}
\usepackage{booktabs}
\usepackage{subcaption}
\usepackage{multirow}
\usepackage{mathrsfs}
\usepackage{threeparttable}
\usepackage{booktabs}

\newcommand{\algname}{SFedSat\xspace}

\def\BibTeX{{\rm B\kern-.05em{\sc i\kern-.025em b}\kern-.08em
    T\kern-.1667em\lower.7ex\hbox{E}\kern-.125emX}}

\begin{document}

\title{A Semi-Supervised Federated Learning Framework with Hierarchical Clustering Aggregation for Heterogeneous Satellite Networks \\

}

\author{Zhuocheng Liu, Zhishu Shen,~\IEEEmembership{Member, IEEE}, Qiushi Zheng,~\IEEEmembership{Member, IEEE}, Tiehua Zhang,~\IEEEmembership{Member,~IEEE}, Zheng Lei, and Jiong~Jin,~\IEEEmembership{Member,~IEEE}
\thanks{Zhuocheng Liu and Zhishu Shen are with School of Computer Science and Artificial Intelligence, Wuhan University of Technology, Wuhan, China (e-mail: u244460865@whut.edu.cn, z\_shen@ieee.org). Zhishu Shen is also with the Hubei Key Laboratory of Transportation Internet of Things, Wuhan University of Technology, Wuhan, China.}
\thanks{Qiushi Zheng, Zheng Lei, and Jiong Jin are with the School of Engineering, Swinburne University of Technology, Melbourne, Australia (e-mail: \{qiushizheng, zlei, jiongjin\}@swin.edu.au)}
\thanks{Tiehua Zhang is with the School of Computer Science and Technology, Tongji University, Shanghai, China (e-mail:  tiehuaz@tongji.edu.cn). He is also with the State Key Laboratory for Novel Software Technology, Nanjing University, China.}
\thanks{This work was supported in part by the National Natural Science Foundation of China (Grant No. 62472332).} 
\thanks{\textit{Corresponding author: Zhishu Shen.}}}

\maketitle

\begin{abstract}
Low Earth Orbit (LEO) satellites are emerging as key components of 6G networks, with many already deployed to support large-scale Earth observation and sensing related tasks. Federated Learning (FL) presents a promising paradigm for enabling distributed intelligence in these resource-constrained and dynamic environments. However, achieving reliable convergence, while minimizing both processing time and energy consumption, remains a substantial challenge, particularly in heterogeneous and partially unlabeled satellite networks. To address this challenge, we propose a novel semi-supervised federated learning framework tailored for LEO satellite networks with hierarchical clustering aggregation. To further reduce communication overhead, we integrate sparsification and adaptive weight quantization techniques. In addition, we divide the FL clustering into two stages: satellite cluster aggregation stage and Ground Stations (GSs) aggregation stage. The supervised learning at GSs guides selected Parameter Server (PS) satellites, which in turn support fully unlabeled satellites during the federated training process. 
Extensive experiments conducted on a satellite network testbed demonstrate that \algname can significantly reduce processing time (up to 3x) and energy consumption (up to 4x) compared to other comparative methods while maintaining model accuracy.
\end{abstract}


\input{1_introduction}
\input{2_relatedwork}
\input{3_model}

\input{4_algorithm}
\input{5_experiment}
\input{6_conclusion}

\bibliographystyle{ieeetr} 
\bibliography{ref}

\vskip -2\baselineskip plus -1fil
\begin{IEEEbiography}[{\includegraphics[width=1in,height=1.25in,clip,keepaspectratio]{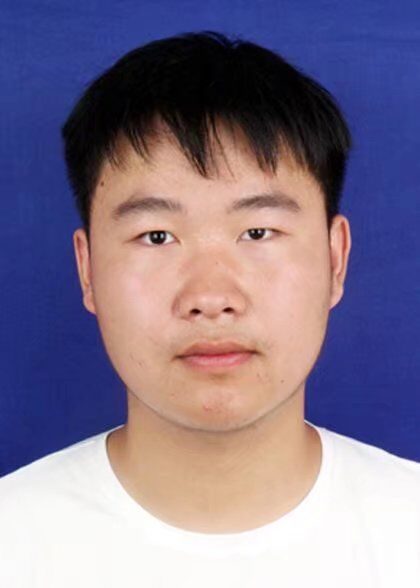}}]{Zuocheng Liu}
received his B.E. degree from the School of
Computer Science and Artificial Intelligence at Wuhan University of Technology, China, in 2023. He is currently working toward a master degree from the School of Computer Science and Artificial Intelligence at Wuhan University of Technology. His major interests include satellite networks and the distributed computing.
\end{IEEEbiography}

\vskip -2\baselineskip plus -1fil

\begin{IEEEbiography}[{\includegraphics[width=1in,height=1.25in,clip,keepaspectratio]{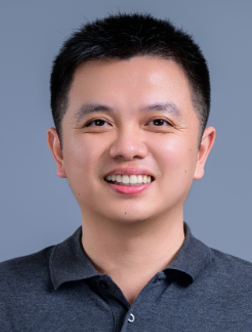}}]{Zhishu Shen}
received the B.E. degree from the School of Information Engineering at the Wuhan University of Technology, China, in 2009, and the M.E. and Ph.D. degrees in Electrical and Electronic Engineering and Computer Science from Nagoya University, Japan, in 2012 and 2015, respectively. He is currently an Associate Professor in the School of Computer Science and Artificial Intelligence, Wuhan University of Technology. From 2016 to 2021, he was a research engineer of KDDI Research, Inc., Japan. His major interests include network design and optimization, edge intelligence, and the Internet of Things.
\end{IEEEbiography}

\vskip -2\baselineskip plus -1fil

\begin{IEEEbiography}
[{\includegraphics[width=1in,height=1.25in,clip,keepaspectratio]{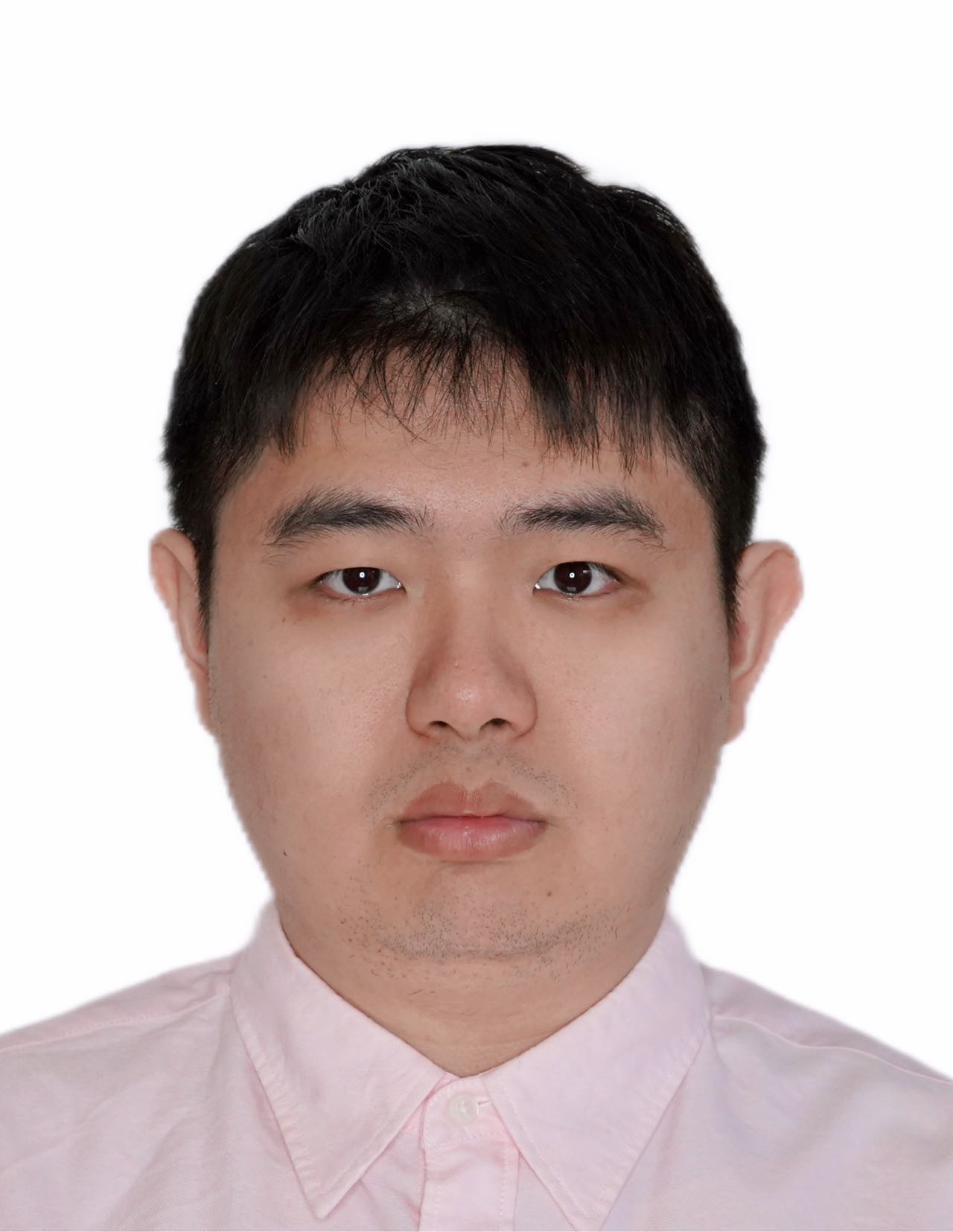}}]{Qiushi Zheng} received the B.E. degree in automation (information and control system) from Beijing Information Science and Technology University, China, in 2015. He received the M.E. degree in engineering science (electrical and electronic) and the Ph.D degree in information communication technology from the Swinburne University of Technology, Australia, in 2017 and 2023. He is currently a Research Fellow at Swinburne University of Technology, Melbourne, Australia. His research interests include satellite-aerial-ground integrated networks, edge computing and the Internet of Things.
\end{IEEEbiography}

\vskip -2\baselineskip plus -1fil

\begin{IEEEbiography}[{\includegraphics[width=1in,height=1.25in,clip,keepaspectratio]{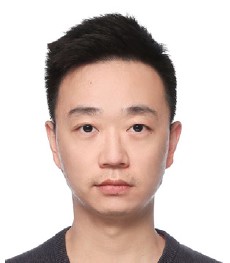}}]{Tiehua Zhang}
received the Ph.D. degree in Computer Science from Swinburne University of Technology, Australia, in 2020. He was a Postdoctoral Researcher with the Department of Computing, Macquarie University from 2020 to 2021, a Research Scientist and team leader at Ant Group from 2021 to 2024. He is currently an Assistant Professor with the School of Computer Science and Technology, Tongji University. His research interests encompass collaborative learning/optimization, edge intelligence, graph learning, and the Internet of Things.
\end{IEEEbiography}

\vskip -2\baselineskip plus -1fil
\begin{IEEEbiography}
[{\includegraphics[width=1in,height=1.25in,clip,keepaspectratio]{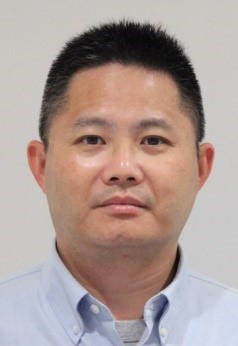}}]{Zheng Lei} is the inaugural Professor of Aviation and Chair of the Department of Aviation at Swinburne University of Technology, Australia. Zheng holds a PhD in Aviation from the University of Surrey in 2006.  He has research interests in aviation business models, aviation market analysis, aviation regulation and policy analysis, and uncrewed aircraft systems. A distinctive feature of his research is its real-world impact and industry focus.  He is also a Board member of Aviation Aerospace Australia. 
\end{IEEEbiography}
\vskip -2\baselineskip plus -1fil

\begin{IEEEbiography}
[{\includegraphics[width=1in,height=1.25in,clip,keepaspectratio]{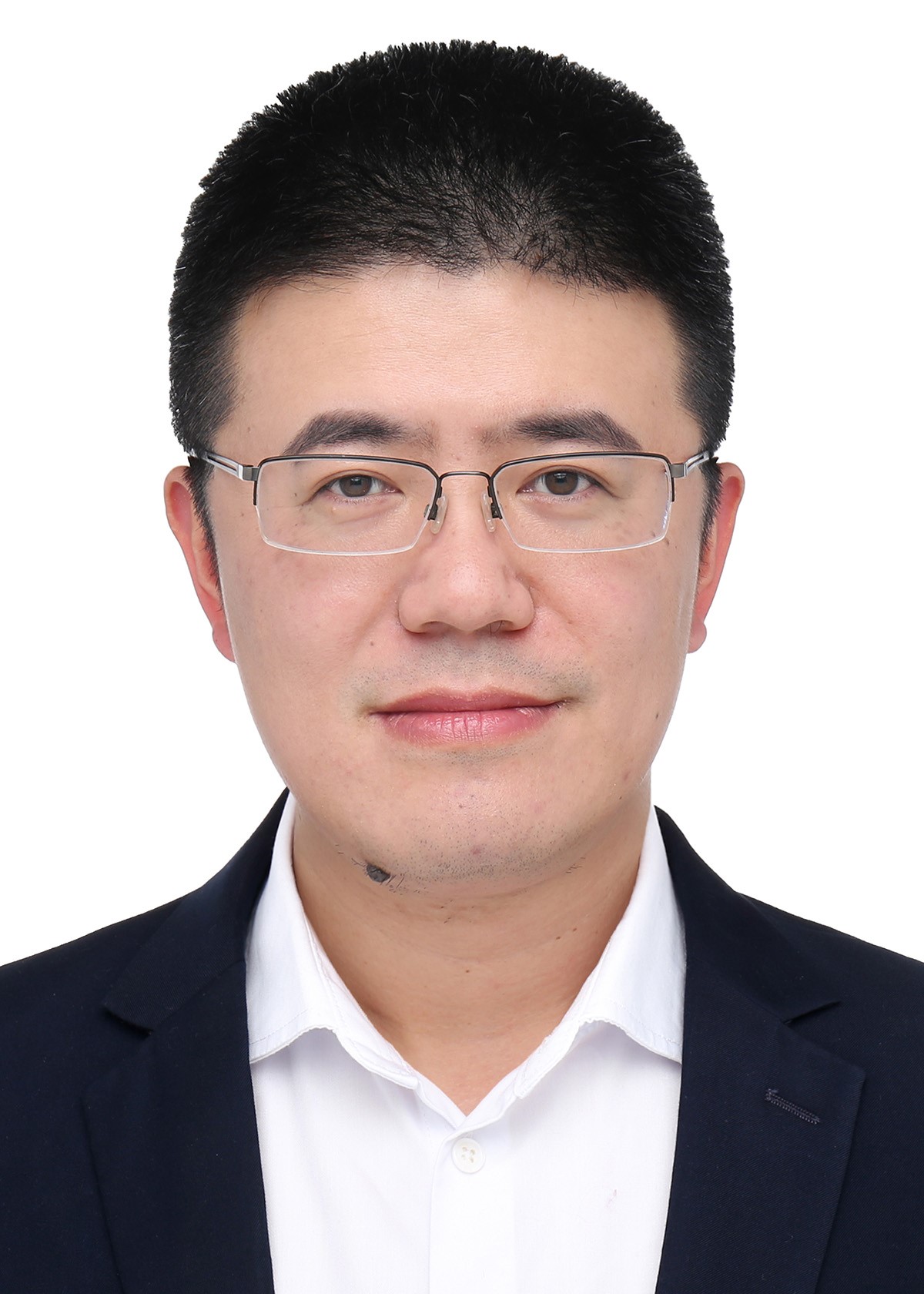}}]{Jiong Jin} received the B.E. degree with First Class Honours in Computer Engineering from Nanyang Technological University, Singapore, in 2006, and the Ph.D. degree in Electrical and Electronic Engineering from the University of Melbourne, Australia, in 2011. He is currently a full Professor in the School of Engineering, Swinburne University of Technology, Melbourne, Australia. His research interests include network design and optimization, edge computing and intelligence, robotics and automation, and cyber-physical systems and Internet of Things as well as their applications in smart manufacturing, smart transportation and smart cities. He was recognized as an Honourable Mention in the AI 2000 Most Influential Scholars List in IoT (2021 and 2022). He is currently an Associate Editor of IEEE Transactions on Industrial Informatics and IEEE Transactions on Network Science and Engineering.
\end{IEEEbiography}

\end{document}

%% file: 1_introduction.tex
\section{Introduction}

To enable high-speed global connectivity and deliver more reliable and secure communication services, Low Earth Orbit (LEO) satellite communication systems are increasingly recognized as a critical component of 6G network architecture~\cite{IEEEMahboob2024}. The rapid expansion of LEO satellite systems is driven by continuous progress in space exploration and a significant decrease in satellite launch and deployment costs. Several large-scale initiatives, such as SpaceX’s Starlink and Amazon’s OneWeb, have been launched to establish expansive satellite networks. These satellite networks provide flexible support for modern Internet applications and are especially valuable in regions lacking traditional terrestrial infrastructure such as oceans and remote rural areas, where they help users overcome geographical constraints and perform computational tasks~\cite{ShenCSUR23,ZhangTMC24,FanTMC25}. Moreover, large satellite constellations are capable of continuously capturing near-real-time remote sensing imagery, thus playing a crucial role in applications like smart agriculture~\cite{LiuSECON23}, environmental monitoring, disaster prevention and response~\cite{ZhangICC24}, and ocean exploration. Collaboration and data sharing among satellites can significantly enhance the accuracy of machine learning models~\cite{PengTNSE25}. However, the large scale of modern LEO constellations and the potentially privacy-sensitive nature of collected data render traditional centralized machine learning approaches, where vast amounts of client data are transmitted to a central cloud server for processing and training, unsuitable for satellite networks.

Federated learning (FL) is a promising distributed learning paradigm with significant potential for large-scale deployment in satellite networks. In FL-based satellite systems, each satellite client performs model training locally and transmits only the model updates to a central server for aggregation after each training round, rather than sending raw data. This approach not only reduces the communication overhead associated with transferring large datasets but also significantly enhances data privacy~\cite{LiaoINFOCOM23,LiTWC24}. The inherent characteristics of FL align well with the unique constraints of satellite networks, offering a cost-effective and privacy-preserving solution for distributed learning in space environments. Therefore, integrating FL into satellite networks is beneficial for enabling practical onboard model training.

Although FL effectively addresses the challenges of limited resources and privacy preservation in satellite networks~\cite{YangTMC24}, several critical issues remain in practical deployment. First, satellite environments are inherently heterogeneous. Satellites vary in orbital configurations, hardware capabilities, communication frequencies, and data distributions, leading to substantial variability across clients~\cite{LinTMC25,ZhaoTMC25}. This heterogeneity can hinder global model aggregation and slow down convergence speed. Specifically, data heterogeneity that implies non-independent and identically distributed data (non-IID data), can lead to divergence of local models, ultimately degrading the performance of the global model~\cite{LiICDE2022}. System heterogeneity further exacerbates the problem, as differences in computational resources lead to varying local training times. As a result, faster clients must wait for slower ones, significantly reducing overall learning efficiency. 
Moreover, in real-world satellite networks, most clients lack the expertise required to annotate data, as they are not domain experts. Additionally, privacy concerns may prevent clients from labeling sensitive information. The continuous generation of large scale data further renders manual annotation infeasible. The majority of existing studies on satellite FL adopt a supervised learning paradigm, assuming that all clients have access to labeled data~\cite{dumeurIEEE24}. This assumption neglects the scarcity of annotations in practical scenarios, resulting in model performance degradation in real-world satellite network environments.



Furthermore, in large-scale satellite constellations, communication overhead grows exponentially. Given the limited communication resources of satellites, it is crucial to minimize the volume of data transmitted~\cite{XiongWCNC24}. In dynamic satellite network environments, communication is restricted to specific daily time windows, and the satellite topology changes dynamically. Therefore, reducing communication time and enabling satellites to complete data exchange as quickly as possible are essential. The key challenge for FL in satellite networks is how to minimize FL processing time and energy consumption while preserving model performance.

To address these challenges, we propose a novel semi-supervised FL framework named \algname, tailored for heterogeneous satellite networks. This framework is designed to improve the efficiency of FL by optimizing both processing time and energy consumption in large-scale satellite constellations. By leveraging labeled data from Ground Stations (GSs) through self-supervised learning, the framework guides semi-supervised learning on satellites to address the challenge of label scarcity in real-world scenarios. The traditional FL process is restructured into two stages: In the satellite cluster aggregation stage, the framework first constructs joint feature vectors by integrating both data characteristics and geographical information of satellite clients. Based on these vectors, the K-means clustering algorithm is applied to group satellites into distinct clusters. Within each cluster, a central satellite client is selected to serve as a Parameter Server (PS), enabling faster and more efficient model aggregation. Each cluster then performs FL respectively, aggregating global model in a decentralized manner. Following this, the process transitions to the GS aggregation stage, where GSs select a subset of cluster PSs with available communication capabilities to further aggregate the global model. Additionally, a model compression mechanism is introduced to dynamically assess the necessity of transmitting quantized global model. The mechanism determines the appropriate quantization bit-width based on gradient variation, effectively balancing communication efficiency and model performance.

The main contributions of this work are summarized as follows:

\begin{itemize}

\item To the best of the authors’ knowledge, this is the first asynchronous, semi-supervised FL framework for heterogeneous satellite networks. In this framework, satellite clients possess entirely unlabeled data, while GS servers hold partially labeled data. Model training alternates between the PS at the GS and the satellite clusters to accelerate convergence and improve aggregation efficiency. In \algname, the optimization of FL processing time and energy consumption is formulated as a joint optimization problem, aiming to maximize model accuracy under resource-constrained satellite environments.
\item We develop a staleness-aware semi-asynchronous aggregation mechanism that selectively incorporates timely client updates to mitigate the effects of system heterogeneity and dynamic connectivity. We also provide theoretical analysis of convergence and computational complexity under non-IID, partially labeled, and resource-constrained conditions specific to satellite environments.

\item We propose a dynamic satellite clustering method based on joint feature vectors, combined with an adaptive sparsification and quantization algorithm to compress model updates based on gradient dynamics. This approach further enhances the system’s adaptability to the dynamic and resource-constrained characteristics of satellite networks.

\item Extensive experiments demonstrate that the proposed framework outperforms state-of-the-art methods on non-IID versions of the CIFAR-10 and the SAT-6 remote sensing datasets. The improvements are consistent across key metrics, including model accuracy, energy consumption, and processing time.

\end{itemize}

The remainder of this paper is organized as follows: Section~\ref{sec:related} reviews the related work. Section~\ref{sec:system_model} presents the system model and formulates the optimization problem based on \algname. Section~\ref{sec:algorithm} summarized the proposed framework \algname. Section~\ref{sec:experiments}
shows the evaluation results that validate the performance of \algname. Finally, Section~\ref{sec:conclusion} concludes this paper.

%% file: 2_relatedwork.tex
\section{Related Work}
\label{sec:related}

\subsection{Federated Learning in Satellite Networks}
Recently, the potential of introducing FL in satellites has been widely discussed because of its ability to enable collaborative training while preserving user privacy. Xu \textit{et al.} designed a connectivity-density-aware FL framework for satellite-terrestrial integration, utilizing ground stations as PSs. The effectiveness of this approach in terms of execution efficiency and accuracy was validated~\cite{XUFGCS24}. Elmahalawy \textit{et al.} used high-altitude platforms as a distributed PS to improve satellite visibility while enhancing FL performance~\cite{ElmahallawySAC24}. Zhou \textit{et al.} proposed an FL framework with partial device involvement. In this framework, the onboard controller strategically selects a portion of devices to upload local model parameters, by which reduces the overall energy consumption on the device side~\cite{ZhouWCNC24}. Chen \textit{et al.} proposed a collaborative satellite FL framework using LEO satellites as PSs, which enables FL to achieve high classification accuracy with relatively low communication latency~\cite{ChenPIMRC23}. Processing the FL global model directly on satellites, rather than offloading parameters to ground stations, is considered promising in terms of scalability and adaptability for large-scale satellite constellations\cite{OhICAIIC25}.

In resource-constrained satellite environments, the need for implementing FL processes in an energy-efficient manner is heightened. Zhou \textit{et al.} designed an FL multi-objective optimization algorithm that leverages decomposition and meta deep reinforcement learning. This algorithm enhances both training efficiency and local training accuracy by addressing multi-objective optimization challenges, resulting in more efficient uploading and aggregation~\cite{ZhouSAC24}. Xia \textit{et al.} proposed a cross domain joint computation offloading algorithm to optimize the trade-off between Age of Information (AoI) and energy consumption when serving multiple traffic categories in satellite networks~\cite{XiaPIMRC23}. Therefore, it is essential to develop a solution that can optimally select PS while reducing the processing time and energy consumption during the FL process.

\subsection{Semi-supervised Federated Learning}
 Semi-supervised FL has emerged to enable collaborative learning from partially labeled decentralized datasets in real-world applications. Chen \textit{et al.} proposed a method to maximize the utilization of unlabeled data by adjusting the optimization objective for high-confidence samples based on the class distributions of both predictions and pseudo-labels\cite{ChenTOE2025}. Jeong \textit{et al.} proposed a federated matching approach that improves upon the naive combination of semi-supervised FL. This method introduces a novel inter-client consistency loss and parameter decoupling mechanism to enable disjoint learning on both labeled and unlabeled data\cite{JeongICLR21}. Diao \textit{et al.} addressed the challenge of combining communication-efficient FL methods such as FedAvg with semi-supervised learning. They proposed an alternative training strategy that utilizes the global model to generate pseudo-labels\cite{DiaoNeurIPS22}. Wang \textit{et al.} introduced an algorithm called FedCPSL, which incorporates adaptive client-side variance reduction, local momentum, and normalized global aggregation to address device heterogeneity and enhance convergence\cite{WangTOW2024}. Liu \textit{et al.} presented a novel approach to improve the performance of semi-supervised FL by leveraging clean knowledge on the server side and unconstrained samples on the client side\cite{LIUNN25}. 
These frameworks leverage semi-supervised learning techniques to automate image annotation and utilize FL to reduce communication overhead. However, most existing semi-supervised FL studies conduct experiments in stable network environments that do not capture the complexity and variability of communication in satellite networks. This may underestimate the impact of satellite orbit movement or communication delay on model synchronization, along with the convergence speed in real-world satellite network environments.

\subsection{Data and Systems Heterogeneity in Federated Learning}
To address the issues caused by data heterogeneity, many studies have focused on enhancing the influence of global information throughout the training process. For example, in HybridFL, a small amount of data shared by clients is used to warm up the global model within a heuristic algorithm, enabling better initial fitting and more stable model updates\cite{YoshidaICC2020}. Wang \textit{et al.} further proposed the use of momentum to accumulate historical aggregation information, as past gradients often contain rich global features that can serve as valuable guidance during model updates\cite{WangIEEEiotj2022}.

To mitigate the impact of system heterogeneity, current mainstream research focuses on staleness-aware weighted aggregation strategies, which adjust each client's contribution to the global model based on the timeliness of its updates. For example, You \textit{et al.} proposed a time-decay-based weighting mechanism that assigns higher weights to more recent updates while reducing the influence of stale ones, thereby improving aggregation quality\cite{YouLOT22}. Similarly, Zhou \textit{et al.} designed a semi-asynchronous aggregation method that prioritizes clients with faster task completion rates for each training round, with the aim of enhancing overall training efficiency and maintaining model freshness\cite{ZhouINFOCOM24}. Building upon existing research on addressing heterogeneity, satellite environments present unique challenges and unexplored areas. A key problem we aim to solve is how to design asynchronous aggregation strategies that are better suited to the dynamic nature of satellite orbits, to more effectively alleviate the impact of heterogeneity.

%% file: 3_model.tex
\section{System Model and Problem Formulation}
\label{sec:system_model}

\subsection{Network Model}

\begin{figure}[tb!]
\centerline{\includegraphics[width=1\linewidth, height=0.6\linewidth]{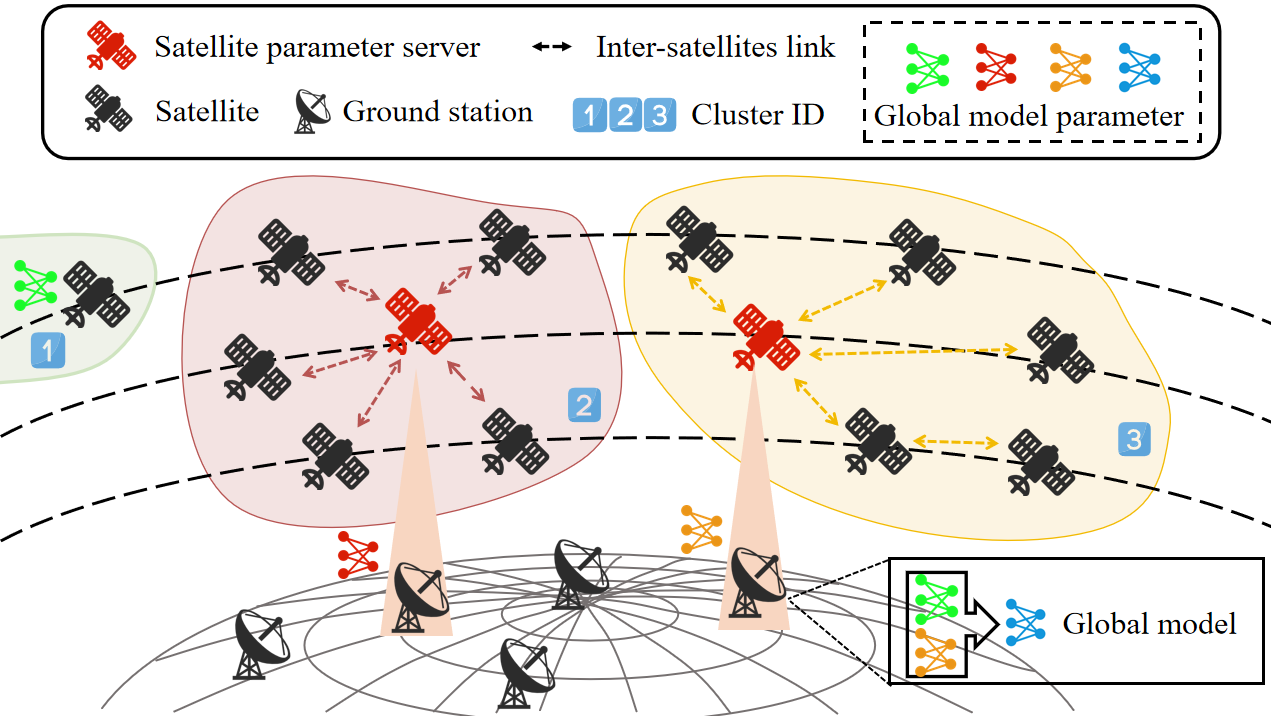}}
\caption{The satellite network model.}
\label{fig:model}
\end{figure}

As illustrated in \figurename~\ref{fig:model}, we consider a satellite communication network consisting of LEO satellites and GSs. To reduce communication overhead between GSs and satellites, selected LEO satellites are designated as PSs. Due to the fixed orbital paths of satellites, the data they collect tends to be highly non-IID. To mitigate the effects of data heterogeneity, the satellite network is partitioned into $K$ clusters using a joint feature vector that combines each satellite client’s geographic location and data characteristics. Each cluster contains $C_k$ satellite clients. In \figurename~\ref{fig:model}, different colors indicate distinct satellite clusters.

Within each cluster, a relatively idle satellite positioned near the geographic center is selected to serve as the local PS. This satellite is responsible for aggregating local model updates from clients within its cluster during the FL process. For global aggregation, each GS $G$ selects the corresponding satellite PSs for collaboration. These selected satellites participate in aggregating the global model. Once global aggregation is completed, the updated models are transmitted back to the respective satellite PSs, completing one full FL cycle.

{\subsection{Satellite Federated Learning}}

The objective of satellite FL is to train a shared global model that enables collaboration among multiple satellite clients without requiring the transmission of private local data to a PS. Specifically, the learning objective is to determine suitable models $w$ to predict the data for input information. We consider a satellite network comprising $C$ clients participating in the FL process, where each client $i \in C$ holds a local dataset $D_i$ of size $|D_i|$. We divide the satellite FL process into two stages. In the satellite cluster aggregation stage,the global loss function over all satellite clients is formulated as:


\begin{equation}
F(w) = \frac{1}{|D|} \sum_{i \in C} \sum_{\Lambda \in D_i} f_i(w; \Lambda)
\end{equation}
where $f_i(w; \Lambda)$ is the loss function calculated by a specific sample $\Lambda$, and $|D|=\sum_{i \in C}|D_i| $ is the total dataset size across all clients. 
The objective of FL is to determine the optimal models $w^*$ that minimizes the global loss function as: 
\begin{equation}
w^* = \arg \min_w F(w)
\end{equation}

In each aggregation round $m$, the satellite PS broadcasts the current global model \(w_m\) to all participating clients. Each client $i$ performs a local gradient descent to update its local model \(w_m^i\). During each FL aggregation round, the clients conduct $e$ epochs of local training. The local gradient is computed using stochastic gradient descent (SGD), formulated as:
\begin{equation}
    \nabla \tilde f_i(w_{m,e}^{i}) = \nabla  \ell(w_{m,e}^{i}; \Lambda)
\end{equation}

For each epoch $e$, the local model \(w_{m,e}^{i}\) evolves as:
\begin{equation}
w_{m,e+1}^{i} = w_{m,e}^{i} - \eta \nabla \tilde{f}_i(w_{m,e}^{i})
\label{eq:evo}
\end{equation}
where $\eta$ represents the learning rate of model. 

After completing the local update, each client transmits its updated models $w_m^i$ to the satellite PS. The satellite PS subsequently aggregates these updates to create the global model for the next round of FL, expressed as:
\begin{equation}
w_{m+1} =  \sum_{i \in C} \frac{|D_i|}{|D|} w_m^i
\label{agg}
\end{equation}

In the ground station aggregation stage, GS collects global models transmitted by its affiliated satellite cluster, performs model aggregation via Equation (\ref{agg}), and subsequently distributes the aggregated global models back to the designated PS.

\vspace{0.1cm}
\subsection{Problem Formulation}

\textbf{Processing time analysis:} At a given time $t_0$, we assume that multiple satellite clusters engage in parallel local training, where each client $i$ holds a local dataset $D_i$. The computing capability (i.e. CPU frequency) of client $i$  is denoted as $f_i$, while $Q$  represents the number of CPU cycles required to train a single data sample. Therefore, the computation time for client $i$ can be expressed as $t_{i}^{cmp}= {D_i Q_i}/{f_i}$.

After completing local training, the client $i$ uploads its local model to the satellite PS for global model aggregation. The achievable transmission rate of client $i$ is given by:

\begin{equation}
r_i = B_i \log_2 \left(1 + \frac{P_0 h_i}{N_0}\right)
\end{equation}
where $N_0$ represents the background noise power, $P_0$ represents transmission power, $B_i$ is transmission bandwidth, and $h_i$ is the channel gain. We assume that the size of the model to be uploaded by each client is a constant $\zeta$, so the communication time for the client $i$ is $t_{i}^{com} = {\zeta}/{r_i}$. We further define $T^m_i$ as the total training time for the $i$-th client in the communication round $m$, which includes both the computation and communication time, formulated as:
\begin{equation}
    T_{i}^{m} =  t_{i}^{\text{cmp}} + t_{i}^{\text{com}} 
\end{equation}

In a semi-asynchronous satellite FL system, we adopt intra-cluster asynchronous aggregation and inter-cluster synchronous aggregation. Each satellite cluster performs asynchronous local aggregation independently, while the GS conducts global aggregation only after all associated clusters have completed their local updates. For a satellite cluster $s_k$, we denote its intra-cluster aggregation time as $T_{s_k}$. It is important to note that the communication and aggregation delay at the GS is primarily determined by the slowest cluster among the set $s_k$ associated with that GS.

In addition to the intra-cluster aggregation time, the total communication time must also include the delay for PS $K_n$ to aggregate the global model and broadcast the updated global model back to its corresponding satellite clusters. This broadcast delay is denoted as $t_{broc}$, and the satellite PS associated with the GS is denoted as $g_{K_n}$.

Accordingly, the total processing time of the FL process can be expressed as:
\begin{equation}
T_c = \sum_{s_k \in g_{K_n}} \left( \max_{i \in s_k}  T_i^m+ T_{s_k} + t_{broc} \right)
\label{eq:time}
\end{equation}

\textbf{Energy consumption analysis}:
The energy consumption in FL can be divided into two parts: the first part is transmission energy consumption, incurred when satellites upload their local model updates to the corresponding PS. After global aggregation is performed at the GS, the updated global model is broadcast back to each satellite client. Accordingly, the transmission energy consumption for satellite model distribution, denoted as $E_{\text{tr}}$, is defined as:
\begin{equation}
E_{\text{tr}} = \sum_{i \in C} P_0 \frac{|w_i|}{r_i}
\end{equation}
where $|w_i|$ is the size of the global model for the $i$-th client. Subsequently, the local training energy \(E_{\text{cmp}}\) in the satellite cluster can be derived as:
\begin{equation}
E_{\text{cmp}} = \sum_{i \in K} \epsilon_0 f^3_i t_i^{\text{cmp}}
\end{equation}
where $\epsilon_0$ is the constant coefficient determined by the hard-
ware architecture. Therefore, the total energy consumption is: 
\begin{equation}
E_c= \min ( E_{\text{tr}} + E_{\text{cmp}}) 
\label{eq:energy}
\end{equation}

\textbf{Objective function}: The objective of this problem is to minimize both processing time $T_c$ (Equation~\ref{eq:time}) and energy consumption  $E_c$ (Equation~\ref{eq:energy}) generated during the FL process. Given the constraints of satellite communication and computational capabilities, the multi-objective optimization function $F(C,K)$ can be formulated as:
\begin{align}
F(C,K) &= \min_{C,K}  \{T_c , E_c\}
\end{align}

The optimization problem for \algname is essentially a discrete combinatorial optimization problem. It involves the selection of PS and the multi-objective optimization of energy consumption and latency, rendering the NP-hard problem~\cite{ZhouWCNC24}. As the number of satellites increases, the solution space expands exponentially, making it computationally infeasible to obtain the global optimum within polynomial time.

%% file: 4_algorithm.tex
\section{Hierarchical Semi-Supervised  Quantitative Federated Learning Framework} 
\label{sec:algorithm}


Due to the NP-hard nature of the satellite optimization problem, formulated as a mixed-integer nonlinear program (MINLP), direct optimization becomes computationally intractable. To address this issue, we decompose the problem into three subproblems and solve each individually. Specifically, we propose an efficient heterogeneous framework based on clustering and two-stage aggregation. This framework decomposes the global optimization problem into subproblems, including satellite cluster partitioning, PS and client selection, and local quantization. This decomposition significantly reduces the overall computational complexity while improving training efficiency and energy consumption in semi-supervised FL, particularly under data and system heterogeneity. Moreover, the proposed framework addresses the critical challenge of effectively handling large-scale unlabeled remote sensing data in heterogeneous satellite networks. In this section, we first provide an overview of the proposed framework, followed by a detailed explanation of its key mechanisms.

\subsection{Overview}
\begin{figure*}[htbp]
\centerline{\includegraphics[scale=0.26]{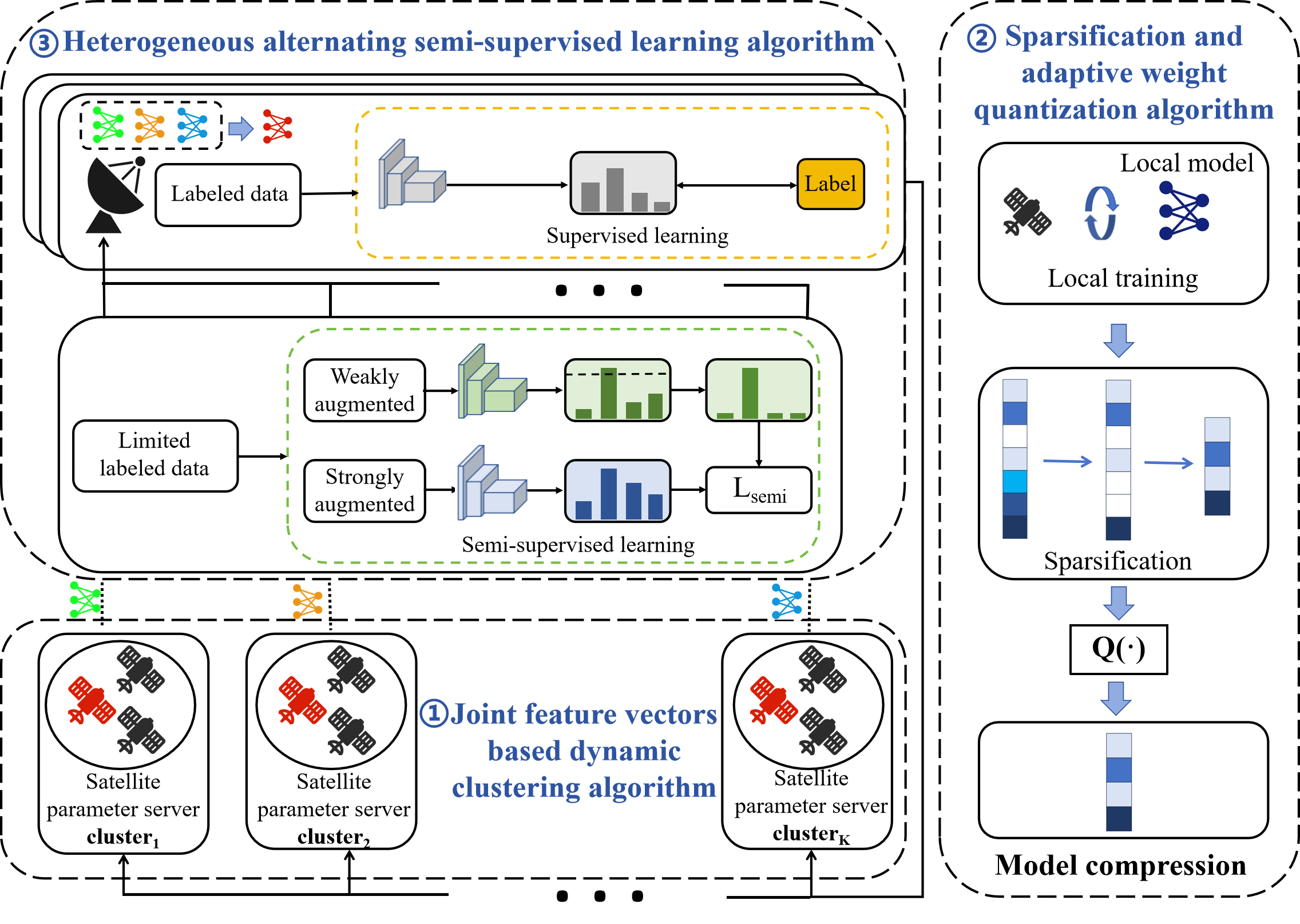}}
\caption{Overview of our proposed framework \algname.}
\label{fig:overview}
\end{figure*}

\figurename~\ref{fig:overview} illustrates the overview of the proposed framework, which consists of three core components: a dynamic satellite clustering mechanism based on joint feature vectors during the satellite clustering phase, a heterogeneous alternating training strategy in the semi-supervised FL process, and a sparsification and adaptive weight quantization algorithm for model transmission.

In the satellite clustering phase, clients with similar system and data characteristics are dynamically clustered based on their joint feature vectors, ensuring that satellites within the same cluster exhibit comparable properties. During the semi-supervised FL process, staleness-aware semi-asynchronous intra-cluster aggregation is performed at the cluster level, while synchronous inter-cluster aggregation is conducted at the central server. To alleviate the limited transmission resources of satellites, we propose a sparsification-based dynamic adaptive quantization algorithm that reduces the bit-width required for transmitting global model. Algorithm~\ref{alg:SemiFL} is introduced in detail across the following three sections.

\begin{algorithm}[ht!]
\caption{Semi-Supervised Heterogeneous Federated Learning for Satellite Networks}
\label{alg:SemiFL}
\begin{algorithmic}[1]
\REQUIRE Satellite info $\mathcal{I}$, clusters $K$, communication rounds $m$, local training epochs $e$, initial confidence threshold $\tau$, beta distribution hyperparameter of CutMix $\mu$, and loss hyperparameter $\lambda$
\ENSURE Global model $w_G$

\STATE Conduct joint feature vectors based dynamic clustering algorithm using $\mathcal{I}$ to form $K$ clusters $\{s_1, \dots, s_k\}$ \label{line:1}
\STATE Select a PS to which GS belongs $s_k$ \label{line:2}

\FOR{each FL round $m$} \label{line:3}
    
    \FOR{each client $i \in s_k$ in parallel} \label{line:4}

        \STATE \COMMENT{Supervised learning at GSs} \label{line:5}
        \STATE Construct supervised dataset $D_s = \{(x, y)\}$ \label{line:6}
        \FOR{each epoch $e$} \label{line:7}
             \STATE Sample data batch $(x_b, y_b)$ of size $A_s$ from $D_s$ \label{line:8}
                \FOR{each batch $(x_b, y_b) \in A_s$} \label{line:9}
                    \STATE Apply weak augmentation: $x_b' = \alpha(x_b)$ \label{line:10}
                    \STATE Compute loss: $L_s = \ell(f(\alpha(x_b), w_i), y_b)$ \label{line:11}
                    \STATE Update model: $w_{i+1} = w_i - \eta \nabla_w L_s$ \label{line:12}
                \ENDFOR \label{line:13}
        \ENDFOR \label{line:14}
        
        \STATE \COMMENT{Semi-supervised training for unlabeled clients} \label{line:15}
        \STATE Generate pseudo-labels with weak augmented $\alpha(x_i)$: ${y}_{i} = f(\alpha(x_{i}), w_{i})$ \label{line:16}
            \STATE Select confident pseudo-labels to form FixMatch set $D^{\text{fix}}_{i} = \left\{(x_{i}, {y}_{i}) \,\middle|\, \max({y}_{i}) \geq \tau \right\}$ \label{line:17}
            
        \FOR{each epoch $e$} \label{line:18}
            \IF{$D^{\text{fix}}_{i} = \emptyset$} \label{line:19}
            \STATE continue \label{line:20}
            \ENDIF \label{line:21}
            \STATE Apply CutMix on $D_i^\text{fix}$ with Beta($\mu, \mu$) to compute $L_{\text{cutmix}}$ \label{line:22}
            \STATE Train with fixMatch and CutMix loss by:
        
                $L_\text{semi} = \lambda L_{\text{fix}} +  (1-\lambda) L_{\text{cutmix}}$
                 \label{line:23}
            \STATE Update model: $w_{i+1} = w_i - \eta \nabla_w L_\text{semi}$ \label{line:24}
        \ENDFOR \label{line:25}
        
    \STATE Conduct sparsification and adaptive weight quantization algorithm to obtain $  Q(S(\Delta w_{i}^m))$\label{line:26} 
    \ENDFOR \label{line:27}
    \STATE \COMMENT{Aggregate satellite cluster models} \label{line:28}
    \FOR{each cluster $k$} \label{line:29}
        \STATE $w_{m+1}^k = w_m^k + \sum_{i \in V_m} \frac{|D_i|}{|D_m|} \cdot p^m_i \cdot
        Q\left(S(\Delta w_{i}^m)\right)$ \label{line:30}
    \ENDFOR \label{line:31}
\ENDFOR \label{line:32}

\STATE \COMMENT{Global aggregation at GS} \label{line:33}
\STATE $w_G = \sum_k \frac{D_k}{D} w_T^k$ \label{line:34}
\RETURN $w_G$ \label{line:35}
\end{algorithmic}
\end{algorithm}

\subsection{Joint Feature Vectors based Dynamic Clustering Algorithm}
The high mobility of satellite clients poses a significant challenge for mitigating data heterogeneity using only data similarity. Clustering clients purely by data characteristics may group clients located in geographically distant regions, resulting in instability as satellites move and dynamically join or leave clusters. To address this issue, we propose a joint feature vectors dynamic clustering algorithm (line~\ref{line:1} in Algorithm~\ref{alg:SemiFL}). The objective is to group satellite clients that share both similar data characteristics and geographical proximity, as such clients are more likely to generate model updates with aligned gradient directions. To evaluate the similarity of data distributions among satellites, we employ cosine similarity between their model update gradients as a proxy metric. The cosine similarity between the gradient updates of two satellites, $C_i$ and $C_j$, can be expressed as:
\begin{equation}
\cos(\vec{C_i}, \vec{C_j}) = \frac{\vec{C_i} \cdot \vec{C_j}}{\|\vec{C_i}\| \cdot \|\vec{C_j}\|}
\end{equation}
where $\vec{C}$ represents the gradient update of the satellite client. Then we normalize the cosine similarity to obtain a data similarity score, defined as:
\begin{equation}
H_{\text{cos}}(i, j) = \frac{1 + \cos(\vec{C}_i, \vec{C}_j)}{2}
\end{equation}

For geographical similarity, we compute the Euclidean distance between the positions of two satellite clients, which is given by:
\begin{equation}
R_{geo}(i, j) = \left\| \mathbf{q}_i - \mathbf{q}_j \right\|_2
\end{equation}
where $\mathbf{q}$ represents the geographic location information of the satellite client. This distance is then normalized to obtain a spatial similarity score, defined as:
\begin{equation}
H_{geo}(i, j) = 1 - \frac{R_{geo}(i, j) - R_{\min}}{R_{\max} - R_{\min}}
\end{equation}
where $R_{\min}$ and $R_{\max}$ denote the minimum and maximum distances among all satellite pairs, respectively. To jointly consider both gradient similarity and spatial proximity, we construct a joint feature vector for each satellite client. For each client $i$, the unified feature vector is defined as:
\begin{equation}
\mathbf{z}_i = \left[ \theta \cdot H_{\text{cos}} \;\middle\|\; (1 - \theta) \cdot H_{\text{geo}} \right]
\end{equation}
where $||$ denotes vector concatenation, and $\theta \in [0, 1]$ is a weighting factor that balances the contribution of gradient and spatial features. These joint feature vectors are then used as inputs to a K-means clustering algorithm, which partitions satellite clients into $K$ groups by minimizing intra-cluster distances in the joint feature space.

\subsection{Heterogeneous Alternating Semi-supervised Learning Algorithm}

For semi-supervised learning in satellite networks, we adopt an alternating training strategy. The process begins with supervised learning on labeled data at the GS (lines~\ref{line:5}–\ref{line:14}). The model is updated by minimizing the standard supervised loss $L_s$ over a batch of labeled data $(x_b, y_b)$ of size $A_s$ randomly sampled from labeled dataset $D_s$, using
\begin{equation}
L_s = \ell(f(\alpha(x_b), w_i), y_b)
\label{eq:supervised_loss}
\end{equation}
where $x$ represents the input image, $y$ denotes its corresponding label, and $\alpha(\cdot)$ denotes a weak data augmentation operation (e.g., random horizontal flipping and cropping). Once the model is updated, GS distributes the trainable parameters $w_i$ to its associated PSs. Upon receiving $w_i$, each PS broadcasts the models to the satellite clients within its cluster (lines \ref{line:15}-\ref{line:25}). Each client then generates pseudo-labels $y_{i}$ as follows:
\begin{equation}
w_{i+1} \leftarrow w_i, \quad {y}_{i} = f(\alpha(x_{i}), w_{i})
\label{eq:pseudo_label}
\end{equation}

Inspired by FixMatch~\cite{fixmatch2020}, each client constructs a high-confidence pseudo-labeled dataset $D^{\text{fix}}_{i}$ at iteration $m$ as:
\begin{equation}
D^{\text{fix}}_{i} = \left\{(x_{i}, {y}_{i}) \,\middle|\, \max({y}_{i}) \geq \tau \right\}
\label{eq:fixmatch}
\end{equation}
where $\tau \in (0,1)$ is a global confidence threshold. If $D^{\text{fix}}_{i} = \emptyset$, the client ceases participation in the current round and does not transmit any updates to the server. For non-empty $D^{\text{fix}}_{i}$, each client performs CutMix-style data augmentation. An equal-size dataset $D^{\text{mix}}_{i}$ is generated via sampling (with replacement) from $D^{\text{fix}}_{i}$. For two randomly sampled pair $(x^1, y^1)$ and $(x^2, y^2)$, CutMix constructs (line \ref{line:22}):
\begin{equation}
x_{\text{cut}} = M \odot x^1 + (1 - M) \odot x^2,\quad 
y_{\text{cut}} = \lambda y^1 + (1 - \lambda) y^2
\label{eq:cutmix}
\end{equation}
where $M$ is a binary mask defining the mixed region, $\odot$ is element-wise multiplication, and $\lambda \sim \text{Beta}(\mu,\mu)$ controls the mixing ratio.

The data batches for FixMatch and CutMix are denoted as $(x_{\text{fix}}, y_{\text{fix}})$ and $(x_{\text{cut}}, y_{\text{cut}})$, respectively. Accordingly, the losses for them are defined as follows:
\begin{equation}
L_{\text{fix}} = \ell(f(\mathcal{A}(x_{\text{fix}}), w_{i}), y_{\text{fix}})
\label{eq:fix_loss}
\end{equation}
\begin{equation}
L_{\text{cutmix}} = \ell\left(f(x_{\text{cut}} , w_i), y_{\text{cut}}\right)
\label{eq:cutmix_loss}
\end{equation}

Here, $\ell$ denotes the cross-entropy loss for classification tasks. $\mathcal{A}$ represents a strong data augmentation, such as RandAugment~\cite{CubukNeurIPS20}, which is employed in our semi-supervised FL framework. Each client updates its local model $w_{i}$ over $e$ local epochs using gradient descent on $L_{\text{cutmix}}$ (line \ref{line:22}).


Through Fixmatch and Cutmix, we can derive the loss function for semi-supervised FL as (line \ref{line:23}):
\begin{equation}
L_\text{semi} = \lambda L_{\text{fix}} +  (1-\lambda) L_{\text{cutmix}}
\label{eq:loss_semi}
\end{equation}

{Therefore, the client performs gradient descent steps with (line \ref{line:24}):
\begin{equation}
w_{i+1} = w_i - \eta \nabla_w L_{semi}
\label{eq:aggregation}
\end{equation}

To mitigate system heterogeneity and account for the dynamic nature of satellites, particularly the possibility that some clients may leave the cluster during the satellite aggregation stage in semi-supervised learning, we propose a staleness-aware aggregation mechanism. This mechanism selects the fastest fraction $\epsilon$ of clients within each cluster to participate in the aggregation. The time threshold $T_s$ is adaptively determined based on the empirical distribution of client completion times from recent training rounds (lines \ref{line:28} to \ref{line:31}). Let $V_m$ denote the ID set of client identifiers corresponding to the fastest fraction  $\epsilon$ of clients in cluster $K$ at iteration $m$, and let their corresponding staleness values be represented by the set $\{\phi^m_i\}_{i \in V_m}$. The staleness of client $i$ at the current global iteration $m$ is defined as: $\phi^m_i = m - m'$, where $m'$ denotes the global iteration round when client $i$ last participated in intra-cluster aggregation.

During the intra-cluster aggregation process, the cluster head $K_n$ assigns a weight $p^m_i$ to each client $i \in V_m$ based on its staleness $\phi^m_i$, following an inverse proportional function:
\begin{equation}
p^m_i = \frac{1}{\phi^m_i}
\label{eq:staleness_weight}
\end{equation}

Subsequently, the staleness-aware semi-asynchronous intra-cluster aggregation performed at each cluster $s_k$ in iteration $m$ is (lines \ref{line:29}-\ref{line:31}):
\begin{equation}
w_{m+1}^k = w_m^k + \sum_{i \in V_m} \frac{|D_i|}{|D_m|} \cdot p^m_i \cdot         Q\left(S(\Delta w_{i}^m)\right)
\label{eq:intra_cluster_aggregation}
\end{equation}

where $|D_m|$ represents the total number of data samples held by clients participating in the intra-cluster aggregation at cluster $K_n$.
Finally, the GS broadcasts the updated global models to all affiliated satellites in their clusters, completing the hierarchically clustered FL process (lines \ref{line:33}-\ref{line:35}).

\subsection{Sparsification and Adaptive Weight Quantization Algorithm}

Due to the spatial and temporal similarity of the data, such as remote sensing data collected by satellites within short time intervals (e.g., consecutive ocean imagery), such data can be more sparsely represented in the transform domain. Meanwhile, in FL, the dynamic range of client gradients varies with aggregation states, necessitating an adaptive sparsification and quantization strategy to balance communication efficiency and model accuracy. To maintain model performance while reducing communication overhead in resource-constrained satellite environments, we adopt an unbiased compression strategy inspired by ~\cite{YanIEEE22} and design a sparsification-based dynamic adaptive quantization algorithm (line \ref{line:26}). This algorithm involves two stages: sparsification of model updates and unbiased adaptive quantization of the non-zero components.

During the sparsification process, we randomly retain $k_{i}^m$ elements from $\Delta w_{i}^m$ and set the remaining entries to zero. The retained update is then scaled by a factor $\delta_{i}^m$:
\begin{equation}
S(\Delta w_{i}^m) = \delta_{i}^m \cdot \text{Rand}_{k_{i}^m}(\Delta w_{i}^m)
\label{eq:sparse}
\end{equation}
where $\text{Rand}_{k_{i}^m}(\cdot)$ selects $k_{i}^m$ coordinates uniformly at random, and the scaling factor is defined as:
\begin{equation}
\delta_{i}^m = \frac{d_w}{k_{i}^m}
\label{eq:scale}
\end{equation}
with $d_w$ being the dimension of $\Delta w_{i}^m$. During the adaptive quantization process, in order to prevent model divergence and degradation in training performance, we adopt an unbiased quantization strategy to ensure that the compressed model remain free from deviation. First, we compute the $\ell_2$ norm of the sparse update:
\begin{equation}
S_{l2} = \| S(\Delta w_{i}^m) \|_2
\label{eq:l2norm}
\end{equation}

Then, for each non-zero element $S(\Delta w_{i}^m)$, we normalize its magnitude:
\begin{equation}
r_j = \frac{ |S(\Delta w_{i}^m)| }{ S_{l2} }
\label{eq:normalize}
\end{equation}

For the $\Delta w_{i}^m$ parameter of each global model vector, we compute its temporal variation compared to the previous round:
\begin{equation}
\Delta_{\text{diff}}(m) = \left| \Delta w_{i}^m - \Delta w_i^{m-1} \right|
\label{eq:variation}
\end{equation}

An adaptive bit-width $b_i^m$ is selected based on the magnitude of change:
\begin{equation}
b_i^m =
\begin{cases}
8, & \text{if } \Delta_{\text{diff}}(m) > Z \\
4, & \text{otherwise}
\end{cases}
\label{eq:adaptive_bit}
\end{equation}
where $Z$ is a predefined threshold controlling the sensitivity.

We define the number of quantization levels as:
\begin{equation}
\varsigma = 2^{b_i^m - 1} - 1
\label{eq:levels}
\end{equation}
where $b_i^m$ represents the number of quantized bits. The interval $[0, S_{l2}]$ is uniformly divided into $\varsigma$ segments. The $k$-th interval is defined as $ M_k = [l_{k-1}, l_k] $, where:
\begin{equation}
\quad l_k = \frac{k}{\varsigma} \cdot S_{l2}, \quad k = 1, \dots, \varsigma
\label{eq:interval}
\end{equation}

For each normalized value $r_j$ within interval $M_k$, we perform stochastic rounding as follows:
\begin{equation}
\zeta(\Delta w_{i}^m, \varsigma) =
\begin{cases}
\frac{k-1}{\varsigma}, & \text{with probability } \frac{l_k - r_j}{l_k - l_{k-1}} \\
\frac{k}{\varsigma},   & \text{with probability } \frac{r_j - l_{k-1}}{l_k - l_{k-1}}
\end{cases}
\label{eq:quant_prob}
\end{equation}

The final quantized value is then reconstructed as:
\begin{equation}
Q(S(\Delta w_{i}^m)) = S_{l2} \cdot \text{sign}(S(\Delta w_{i}^m)) \cdot \zeta(\Delta w_{i}^m, \varsigma)
\label{eq:quantized}
\end{equation}

This quantization process is unbiased, satisfying:
\begin{equation}
\mathbb{E} \left[ Q(S(\Delta w_{i}^m)) \right] = S(\Delta w_{i}^m)
\label{eq:unbiased}
\end{equation}

\subsection{Convergence Analysis} 

We analyze the convergence of the proposed algorithm based on standard assumptions from federated optimization~\cite{LiICLR20} as follows:

\textbf{Assumption 1 (Smoothness).} Each local objective $F_i(w)$ is $L$-smooth, i.e.,
\begin{equation}
\|\nabla F_i(w) - \nabla F_i(w')\| \leq L_{smooth} \|w - w'\|, \quad \forall w, w'
\end{equation}
where $L_{smooth}$ is a positive constant.

\textbf{Assumption 2 (Bounded Variance).} The variance of local stochastic gradients is bounded:
\begin{equation}
\mathbb{E}_{\xi_i}[\|\nabla \ell(w; \xi_i) - \nabla F_i(w)\|^2] \leq \sigma^2
\end{equation}
where ${\xi_i}$ denotes the samples drawn from the local dataset of the client, and $\sigma^2$ represents the upper bound on the variance of the stochastic gradient.

\textbf{Assumption  3 (Client Drift).} The divergence between local and global gradients is bounded:
\begin{equation}
\mathbb{E}_i[\|\nabla F_i(w) - \nabla F(w)\|^2] \leq \iota^2
\end{equation}
where $\iota^2$ represents the upper bound of Client Drift between clients.

\textbf{Assumption  4 (Unbiased Compression).} The gradient compression operator $Q(\cdot)$ satisfies:
\begin{equation}
\quad \mathbb{E}[\|Q(S(\Delta w_i)) - \Delta w_i\|^2] \leq \omega \|\Delta w_i\|^2
\end{equation}
where $\omega$ represents the upper bound of the relative variance of the compression error term.

For theoretical convergence analysis, we adopt a simplified synchronous model update of the form:
\begin{equation}
w_{m+1} = w_m - \eta \cdot \frac{1}{|V_m|} \sum_{i \in V_m} Q(S(\Delta w_{i}^m))
\end{equation}

Define the compression noise for client $i$ at round $m$ as:
\begin{equation}
\mu_i{(m)} = Q(S(\Delta w_{i}^m)) - \Delta w_{i}^m
\end{equation}

Since $\mathbb{E}[\mu_i{(m)}] = 0$, we have:
\begin{align}
\mathbb{E}[Q(S(\Delta w_{i}^m))] &= \mathbb{E}\left[\frac{1}{|V_m|}\sum\limits_{i \in V_m} \Delta w_{i}^m\right] = \bar{\Delta}_m
\end{align}
where $\mathbb{E}[\mu_i{(m)}] = 0$ and $\mathbb{E}[\|\mu_i{(m)}\|^2] \leq \omega \|\Delta w_{i}^m\|^2$ by Assumption  4. Using the bias and compression error assumptions for second-order moment control:
\begin{equation}
\mathbb{E}[\|Q(S(\Delta w_{i}^m))\|^2] \leq 2\mathbb{E}[\|\bar{\Delta}_m\|^2] + 2\mathbb{E}[\|\bar{\mu}_m\|^2] \leq 2\iota^2 + 2\omega
\end{equation}

According to Assumption  1, we use Taylor expansion to write the decrease in objective function values before and after each round of optimization as:
\begin{align}
F(w_{m+1}) &\leq F(w_m) + \langle \nabla F(w_m), w_{m+1} - w_m \rangle \nonumber \\
    &\quad + \frac{L_{smooth}}{2} \|w_{m+1} - w_m\|^2
\label{eq:smooth}
\end{align}

Substituting this into~\eqref{eq:smooth} and taking expectation, we obtain:

\begin{align}
\mathbb{E}[F(w_{m+1})] &\leq \mathbb{E}[F(w_m)] \notag \\
&\quad - \eta \cdot \mathbb{E}[\langle \nabla F(w^{(m)}), \bar{\Delta}_m \rangle] \notag \\
&\quad + \frac{L_{smooth}\eta^2}{2}(2\iota^2 + 2\omega)
\label{equ_1}
\end{align}

Using Young’s inequality and Assumption 3:
\begin{align}
\langle \nabla F(w), \bar{\Delta}_m \rangle &\geq \frac{1}{2}\|\nabla F(w)\|^2 - \frac{1}{2}\|\bar{\Delta}_m - \nabla F(w)\|^2 \notag \\
&\quad \geq \frac{1}{2}\|\nabla F(w)\|^2 - \iota^2
\end{align}

Plug back into~\eqref{equ_1}:
\begin{align}
\mathbb{E}[F(w_{m+1})] \leq &\mathbb{E}[F(w_m)] - \frac{\eta}{2}\mathbb{E}[\|\nabla F(w_m)\|^2] \notag \\
&\quad + \eta\iota^2 + L_{smooth}\eta^2(\iota^2 + \omega)
\end{align}

Sum from iteration $m$:
\begin{align}
\frac{1}{m} \sum_{m=0}^{m-1} \mathbb{E}[\|\nabla F(w_m)\|^2] &\leq \frac{2(F(w_0) - F(w_C)}{\eta m} \notag \\
&\quad + 2\iota^2 + 2L_{smooth}\eta(\iota^2 + \omega) \notag \\
&\quad \leq \mathcal{O} \left( \frac{1}{\eta {m}} + {\iota^2} + \lambda \varepsilon_{\text{ssl}} + \omega \right)
\end{align}
where $\varepsilon_{\text{ssl}}$ denotes the unsupervised estimation error.

The above proof indicates that our \algname can ultimately converge and the error term is controllable.

\subsection{Time Complexity Analysis}
We analyze the time complexity of the proposed framework algorithm over $m$ communication rounds. The overall time complexity is primarily composed of the following components:

\begin{itemize}

    \item \textbf{Joint feature vectors feature based dynamic clustering algorithm}: The computation of the joint feature vector for each client has a complexity of $O(d_gC^2)$, where $d_g$ is the dimension of the gradient vector. Under $m_{\text{k}}$ iterations of the K-means algorithm, the clustering complexity is $O(m_{\text{k}}KC^2)$. Therefore, the overall time complexity of the joint feature based dynamic clustering algorithm is $O(d_g \cdot C^2 + m_{\text{k}} \cdot K \cdot C^2)$.

    \item \textbf{Heterogeneous alternating semi-supervised learning algorithm}: Each client performs $e$ epochs of local training per round. Let $\bar{D}$ be the average number of samples per client, $E_{cmp}$ the computational energy consumption per sample, and $C$ the number of clients. The total time complexity for local training is $\mathcal{O}(m \cdot C \cdot e \cdot \bar{D} \cdot E_{cmp})$.

    \item \textbf{Sparsification and adaptive weight quantization algorithm}: The overall sparsification complexity is $O(d_w)$, and the total quantization complexity is $O(k_{i}^m)$. Since $k_{i}^m \ll d_w$, this incurs a time complexity of $O(m \cdot C \cdot d_w)$.

\end{itemize}

Therefore, the time complexity of the entire training process is $\mathcal{O}\left( 
    (d_g + m_{\text{k}} \cdot K) \cdot C^2
    + m \cdot C \cdot (e \cdot \bar{D} \cdot E_{\text{cmp}} + d_w)
\right)$.


%% file: 5_experiment.tex
\section{Experiment}
\label{sec:experiments}

\subsection{Experimental Setup}
We develop a simulated satellite network testbed and conduct extensive experiments to evaluate the effectiveness of the proposed framework. The experiments are conducted on a CPU-based server equipped with 32 GB of 4800-MHz DDR5 memory and an Intel Core i7-13700KF processor running at 3.40 GHz.

Our proposal is validated on simulated networks of two different scales: 100 and 3000 satellites. These LEO satellites are uniformly distributed within each orbital plane. Each client performs training using mini-batch SGD with a batch size of 64 and an initial learning rate of 0.01. We employ two datasets for evaluation: CIFAR-10 and SAT-6~\cite{SAT6}. Due to the limited size of the CIFAR-10 dataset, which hinders convergence in large-scale satellite FL scenarios, the SAT-6 remote sensing image dataset is used for experiments involving 3000 satellites. For CIFAR-10, we adopt Wide ResNet-28$\times$2~\cite{Resnet} as the training model, while for SAT-6, we utilize a shallow CNN. The shallow CNN consists of three convolutional layers, each composed of a 3$\times$3 convolutional layer (stride = 1, activation = ReLU), followed by a 4$\times$4 max pooling layer (stride = 2), and is followed by a fully connected layer. Additional main parameters are summarized in Table~\ref{tab:params}.

To simulate non-IID data distributions, 20\% of the samples in each dataset were assigned to a specific class for each client, while the remaining 80\% were randomly drawn from the other classes. In the semi-supervised FL setting, labeled data constitutes 10\% of the total dataset, and all labeled data is located at GSs. In our composite-feature-based dynamic clustering algorithm, the clustering threshold $\theta$ was set to 0.4. After partitioning the satellite clients into $K$ clusters, each client performs local training for 500 rounds. Subsequently, GSs aggregated the model updates from their respective satellite clusters. For adaptive gradient quantization, we initialize the quantization level to 8 bits, following the method proposed in~\cite{AlistarhNIPS2017}. However, global models with gradient magnitudes below 0.01 were quantized to 4 bits to reduce communication consumption. 

For baseline comparisons, hyperparameters were configured according to the settings recommended in the corresponding original studies. Once each algorithm achieved the target accuracy, we evaluated its performance based on the total wall-clock training time, which includes computation time, communication time, and all associated overheads.

\begin{table}[tb]
 \caption{Main experimental parameters.}
 \label{tab:params}
 \centering
 {
\begin{tabular}{ll}
   \hline
   Parameter  & Value \\
   \hline \hline
   Cluster $K$ (region) &  3-7 (CIFAR-10), 20-60 (SAT-6) \\
   Client $C$ (satellite) & 100 (CIFAR-10), 3000 (SAT-6)\\
   Ground station number & 2\\
   Satellite altitude & 1300~km\\
   Satellite inclination & 53 degree\\
   Quantization level & 8-bit, 4-bit\\
   Client selection rate $\epsilon$ & 0.6\\
   Gradient change threshold $Z$ & 0.01\\
   CPU cycle $f_i$ & 50 GC/s (GC = $10^9$ cycles)~\cite{ZhuJSAC23}\\
   Transmission bandwidth $B_i$ & 27~GHz (Ka-band)\\
   Satellite transmission power $p_i$ & 30~dBW~\cite{ZhangIoT23} \\
   Noise power density $N_0$& -174~dBm/Hz~\cite{ZhuJSAC23} \\
   Batch size & 64\\
   Momentum & 0.9\\
   \hline
  \end{tabular}
  }
\end{table}

\subsection{Comparative Methods}
We evaluate the effectiveness of our proposed Semi-supervised FL in Heterogeneous satellite networks (\textbf{\algname}) against the following four comparative methods.
\begin{itemize}

    \item\textbf{C-FedAvg}~\cite{ChenPIMRC23}: A centralized FL method based on FedAvg, where clients send their data to a central satellite server for model training.
    

    \item\textbf{H-BASE}~\cite{LiuICC20}: A clustered FL method that performs training through a fixed number of intra-cluster aggregation iterations, using randomly selected clients within each cluster.
    
    \item\textbf{FedMatch}\cite{JeongICLR21}: A semi-supervised FL method that introduces a novel inter-client consistency loss and a parameter decoupling strategy to enable disjoint learning on both labeled and unlabeled data.
    \item\textbf{SemiFL}\cite{DiaoNeurIPS22}: A semi-supervised FL method in which the server fine-tunes the global model using labeled data, while clients perform semi-supervised learning by generating pseudo-labels with the global model.

\end{itemize}

\subsection{Experimental Results}

\begin{figure}[tb!]
	\centering
	\begin{minipage}[b]{.493\columnwidth}
		\centering
		\includegraphics[width=\columnwidth]{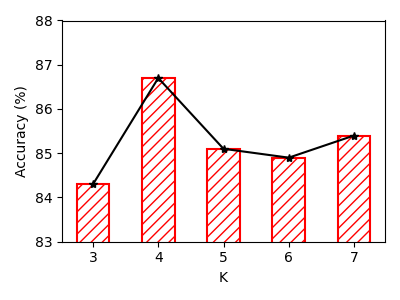}
		\subcaption{CIFAR-10}\label{fig:1_cluster}
	\end{minipage}
        \begin{minipage}[b]{.493\columnwidth}
		\centering
		
		\includegraphics[width=\columnwidth]{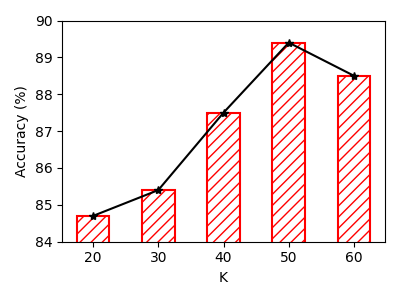}
		\subcaption{SAT-6}\label{fig:2_cluster}
	\end{minipage}

	\caption{Impact of parameter $K$}
    
	\label{fig:cluster}
\end{figure}

\textbf{Impact of clustering algorithm}: To evaluate the impact of our proposed clustering algorithm, we experimented with different values of $K$ to examine how the number of clusters influences model performance. \figurename~\ref{fig:cluster} illustrates the global model accuracy as a function of $K$.

For both the CIFAR-10 and SAT-6 datasets, we observe that increasing $K$ initially leads to improved global model accuracy, which then gradually plateaus. This trend can be attributed to the following: when $K$ is small, each cluster contains a larger number of clients, resulting in greater intra-cluster heterogeneity. Such diversity in data distributions and system characteristics makes it difficult for the global model to learn effectively. As $K$ increases, clients within each cluster become more homogeneous, enabling more consistent and effective local training, which in turn enhances global model performance. However, beyond a certain point, further increasing $K$ yields diminishing returns.

Considering that a higher number of clusters imposes additional communication overhead between satellites and the server, we set $K = 6$ for CIFAR-10 and $K = 50$ for SAT-6 in the subsequent experiments to achieve a balance between performance and communication efficiency.

\textbf{Training convergence}: To validate the effectiveness of our proposed framework, we compare the global model accuracy and training loss over 500 communication rounds against four baselines on different datasets. This evaluation aims to demonstrate both the accuracy and convergence speed of our approach, highlighting its advantages and distinguishing characteristics relative to these alternatives.

\begin{figure}[tb!]
	\centering
	\begin{minipage}[b]{.493\columnwidth}
		\centering
		\includegraphics[width=\columnwidth]{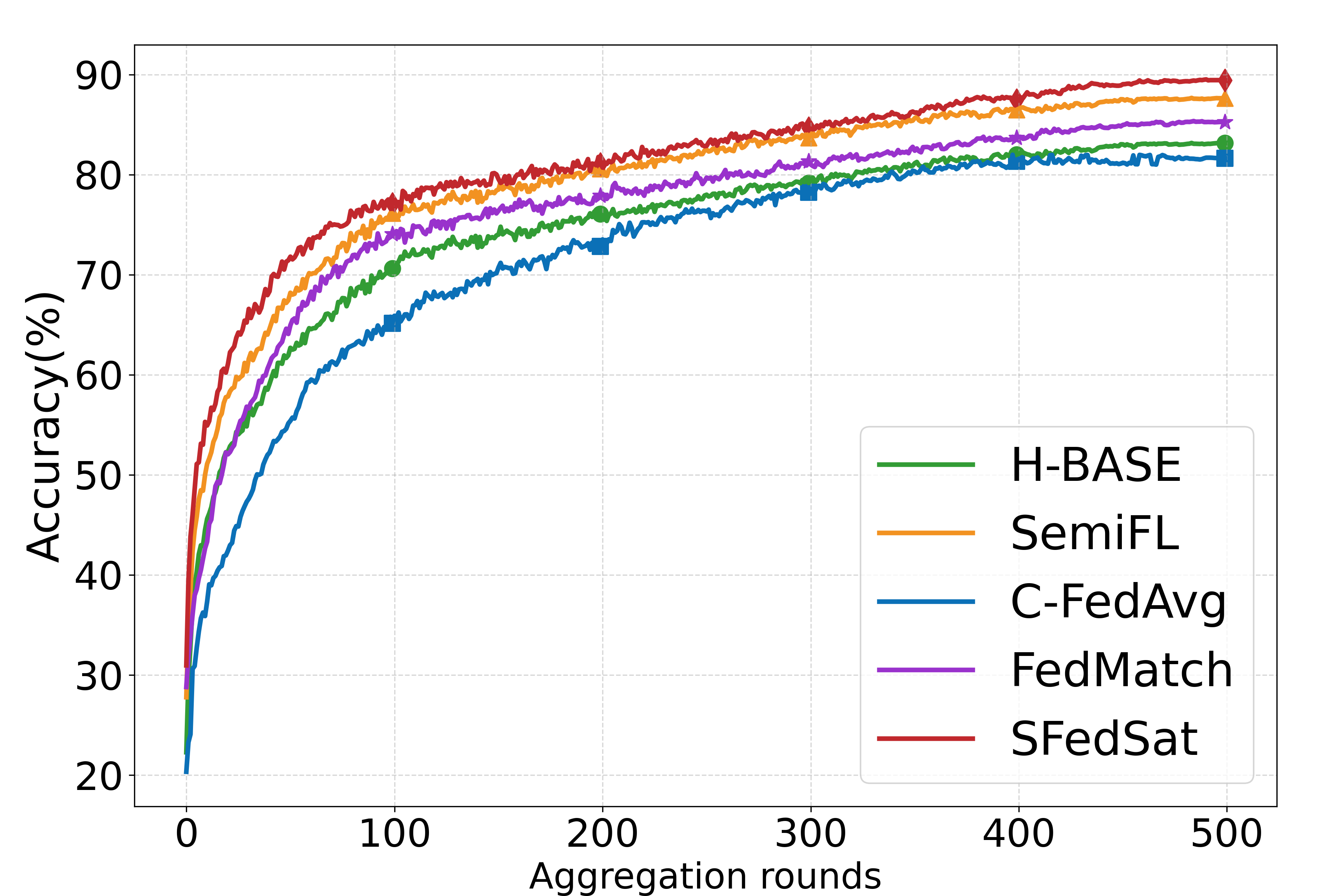}
		\subcaption{CIFAR-10}\label{fig:1_acc}
	\end{minipage}
        \begin{minipage}[b]{.493\columnwidth}
		\centering
		
		\includegraphics[width=\columnwidth]{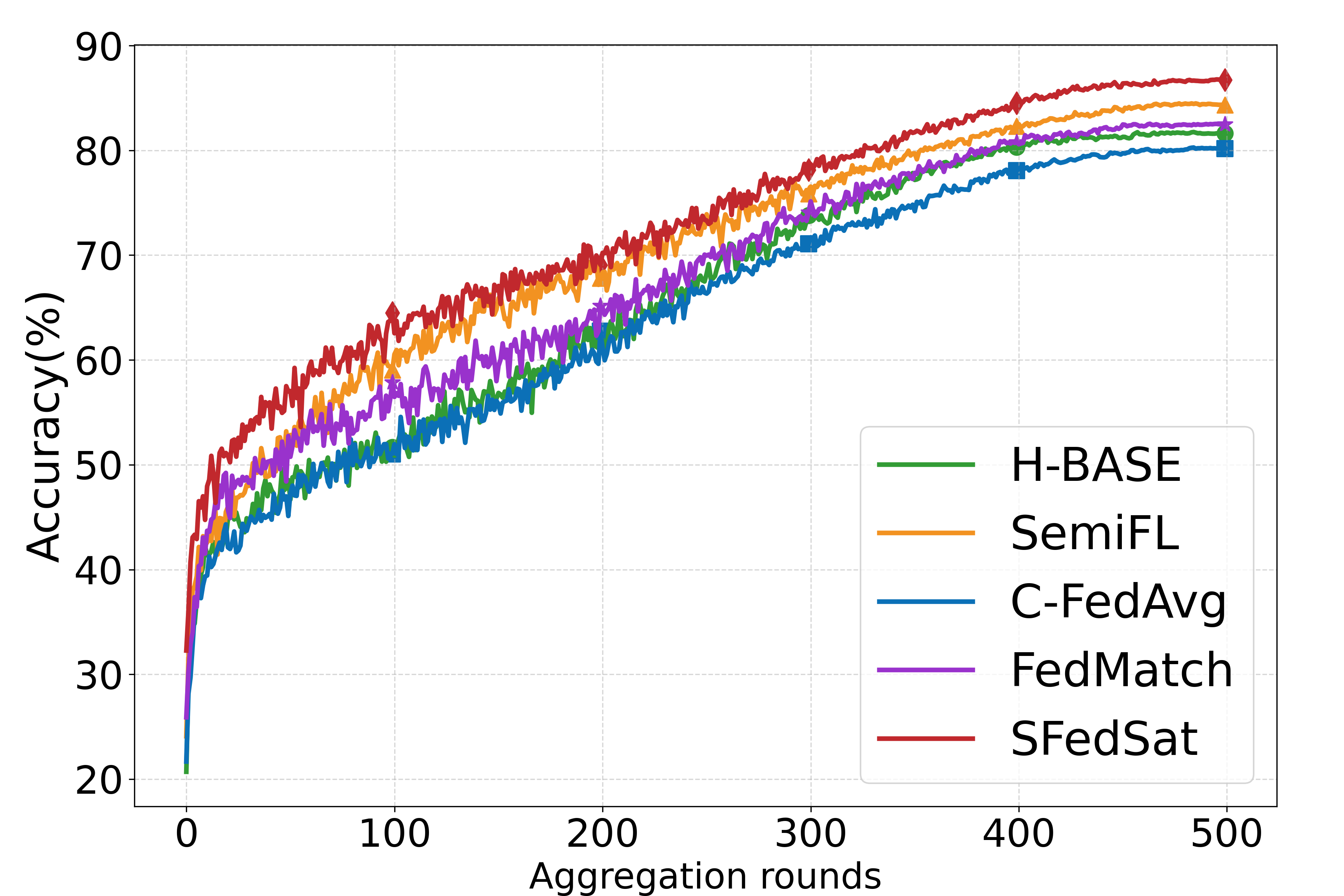}
		\subcaption{SAT-6}\label{fig:2_acc}
	\end{minipage}

	\caption{Accuracy performance for different methods.}
    
	\label{fig:1}
\end{figure}

\begin{figure}[tb!]
	\centering
	\begin{minipage}[b]{.493\columnwidth}
		\centering
		\includegraphics[width=\columnwidth]{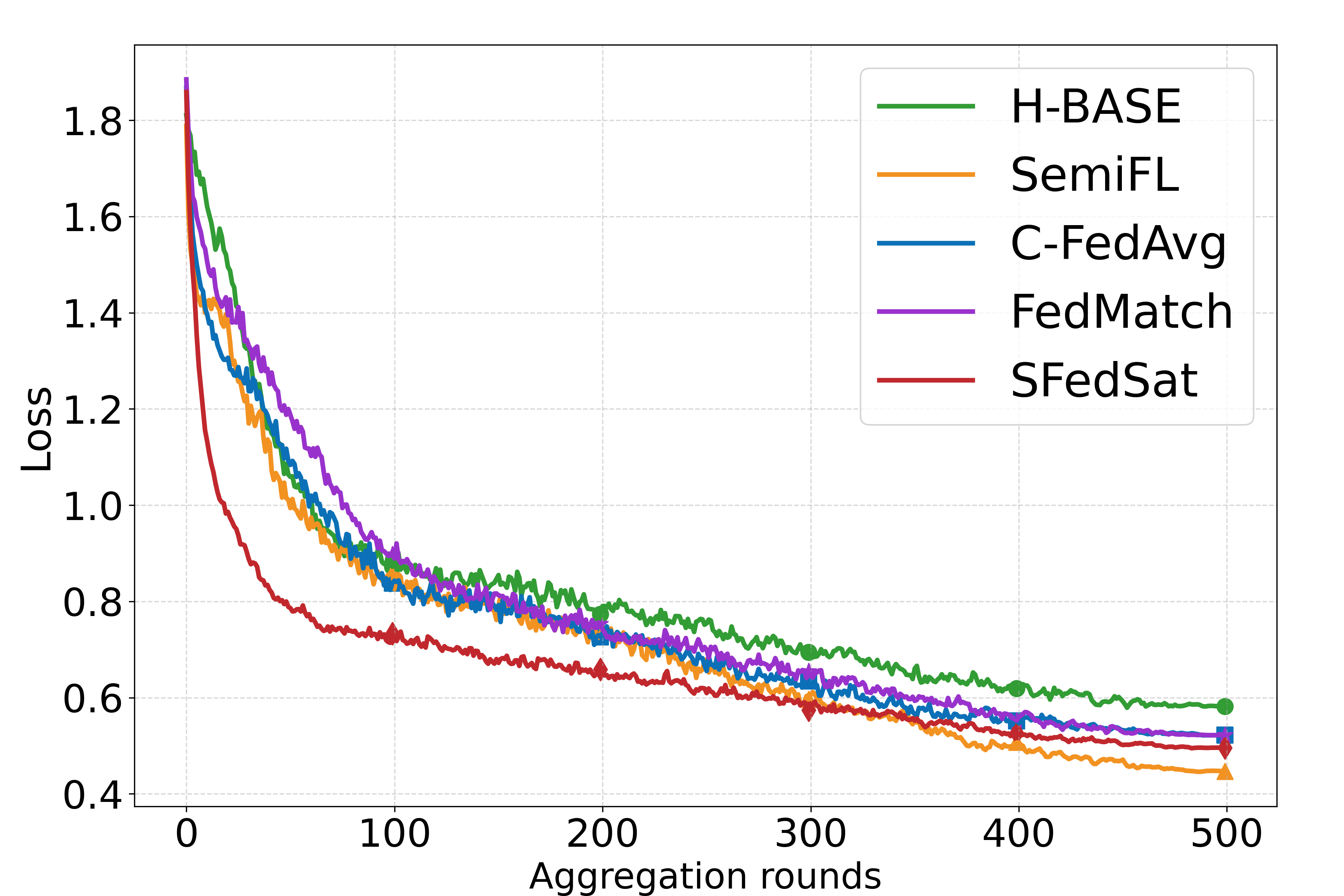}
		\subcaption{CIFAR-10}\label{fig:1_loss}
	\end{minipage}
        \begin{minipage}[b]{.493\columnwidth}
		\centering
		
		\includegraphics[width=\columnwidth]{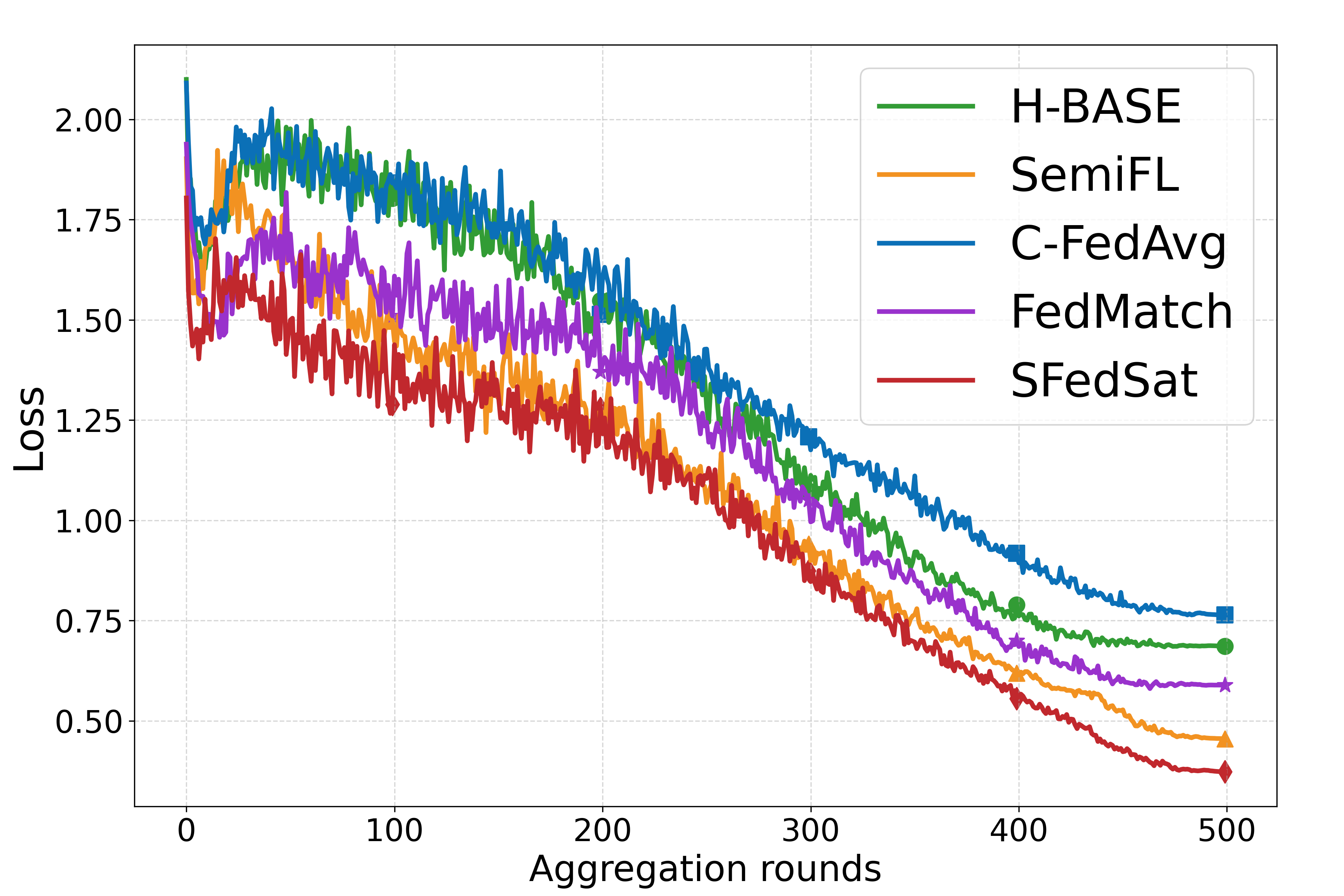}
		\subcaption{SAT-6}\label{fig:2_loss}
	\end{minipage}

	\caption{Loss performance for different methods.}
    
	\label{fig:2}
\end{figure}

Figs.~\ref{fig:1}-\ref{fig:2} show the training loss and model accuracy of various methods under different clustering configurations. Across multiple datasets, our proposed framework achieves relatively rapid convergence and a significant reduction in training loss. This improvement is primarily attributed to the use of joint feature vectors for clustering similar satellite clients, which effectively mitigates the challenges posed by non-IID data distributions. Additionally, an alternating training strategy is employed, where GSs guide the training of satellite client clusters, further enhancing convergence efficiency.

After 500 communication rounds, the accuracy of our framework generally surpasses that of other baseline methods. Despite compressing the transmitted global models during the uplink using \algname, the framework still maintains competitive accuracy. These results demonstrate the feasibility of applying sparsification and quantization to reduce global models in large-scale satellite networks without substantially degrading model performance.

\begin{figure}[tb!]
	\centering
	\begin{minipage}[b]{.493\columnwidth}
		\centering
		\includegraphics[width=\columnwidth]{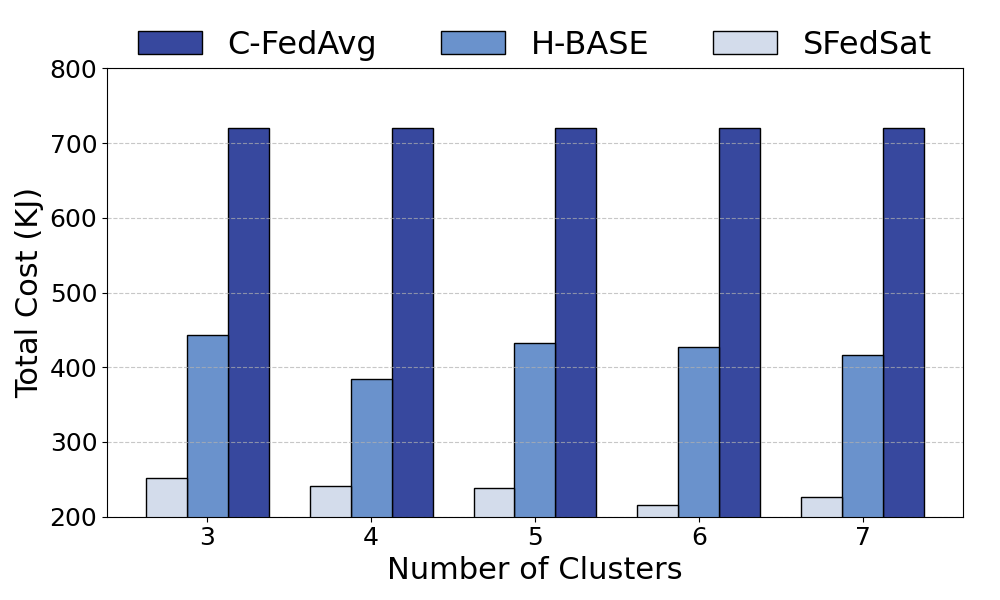}
		\subcaption{CIFAR-10 }\label{fig:1_cost}
	\end{minipage}
        \begin{minipage}[b]{.493\columnwidth}
		\centering
		
		\includegraphics[width=\columnwidth]{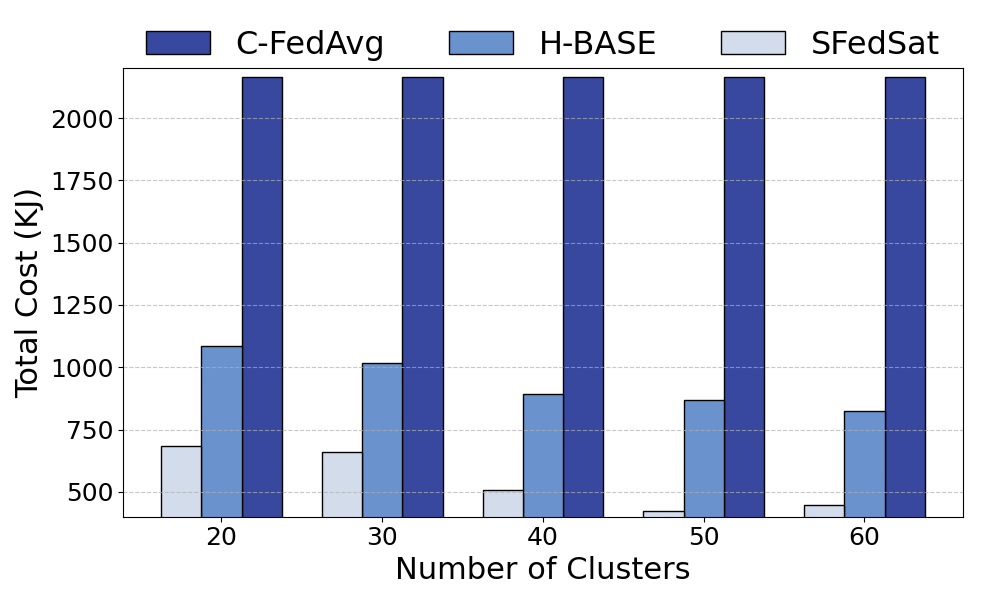}
		\subcaption{SAT-6}\label{fig:2_cost}
	\end{minipage}

	\caption{Total energy consumption for different methods.}
    
	\label{fig:cost}
\end{figure}

\begin{figure}[tb!]
	\centering
	\begin{minipage}[b]{.493\columnwidth}
		\centering
		\includegraphics[width=\columnwidth]{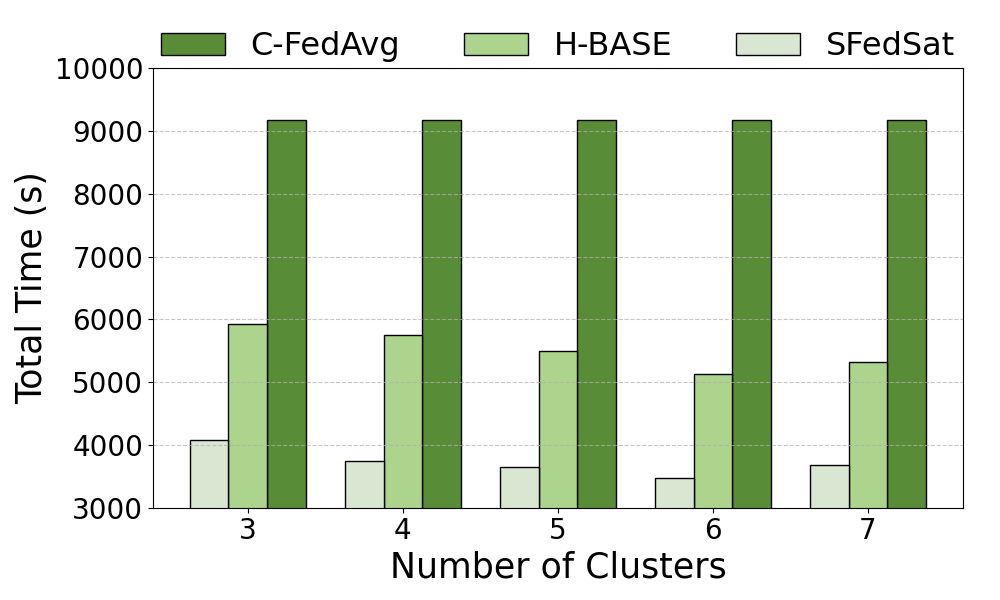}
		\subcaption{CIFAR-10}\label{fig:1_time}
	\end{minipage}
        \begin{minipage}[b]{.493\columnwidth}
		\centering
		
		\includegraphics[width=\columnwidth]{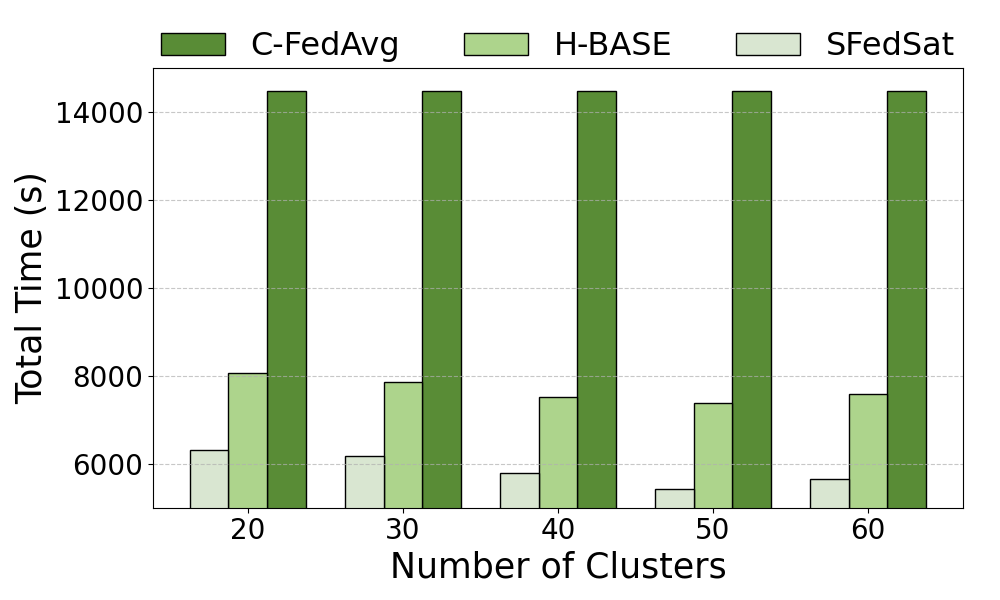}
		\subcaption{SAT-6}\label{fig:2_time}
	\end{minipage}

	\caption{Processing time for different methods.}
    
	\label{fig:time}
\end{figure}

\textbf{FL processing time and energy consumption}: In subsequent experiments, all algorithms are evaluated under the condition of reaching a target accuracy of 80\%. The FL processing time includes the computation time and communication time of the satellites, as defined in Equation~\ref{eq:time}. The total energy consumption of satellites encompasses both the transmission energy consumption and the energy consumption for global model aggregation, as described in Equation~\ref{eq:energy}. \figurename~\ref{fig:cost} and \figurename~\ref{fig:time} show the total energy consumption and processing time across different datasets.

As shown in \figurename~\ref{fig:cost}, our proposed framework \algname achieves substantial energy savings across both datasets, exceeding 3$\times$ reduction compared to C-FedAvg and approximately 2$\times$ compared to H-BASE. Specifically, our method consumes only about one-fifth of the energy required by C-FedAvg in the SAT-6 dataset with $K = 60$. These results demonstrate that our method achieves superior energy efficiency compared to the other two FL baselines. This improvement is attributed to our hierarchical clustered FL framework, which organizes satellite clients into layered clusters, thereby reducing long-range communication with the central server and lowering overall transmission energy consumption. In addition, the adoption of model compression strategies significantly decreases uplink transmission energy consumption. These advantages make our proposal particularly beneficial for large-scale resource-constrained satellite networks.

As shown in \figurename~\ref{fig:time}, under different values of clustering parameter $K$, the clustering-based algorithm achieves approximately 2$\times$ speedup in FL processing time compared to centralized methods. This efficiency gain is attributed to the local local satellite cluster aggregation, which shortens synchronization cycles and thereby accelerates the overall FL communication process. Furthermore, since synchronous methods such as C-FedAvg and H-BASE suffer from straggler effects that delay overall progress, our proposed \algname demonstrates the best performance. Specifically, in both datasets, the completion time of C-FedAvg is 2$\times$ H-BASE's and 3$\times$ \algname's.

\begin{table}[]
\caption{Impact of quantization.}
\label{tab:quantization}

\begin{threeparttable}
\scalebox{0.85}{
\begin{tabular}{c|c|cc|c}
\hline
\multirow{2}{*}{C} & \multirow{2}{*}{K} & \multicolumn{2}{c|}{Weight size (MB) before/after quantization\tnote{1}} & \multirow{2}{*}{Reduction} \\ \cline{3-4}
                   &                    & Before                          & After                          &                            \\ \hline
\multirow{5}{*}{100 (CIFAR-10)} & 3                & 7300                     & 1048.85                     & -6.96x                   \\
                   & 4               & 7300                     & 994.55                    & -7.04x                   \\
                   & 5               & 7300                     & 1015.30                    & -7.19x                   \\
                   & 6               & 7300                     & 1022.41                    & -7.34x                   \\
                   & 7               & 7300                     & 1005.51                    & -7.26x                   \\ \hline

\multirow{5}{*}{3000 (SAT-6)} & 20                & 1210
                     & 234.04                    & -5.17x                   \\
                   & 30               & 1210
                     & 224.91                    & -5.38x                   \\
                   & 40               & 1210
                     & 219.60                    & -5.51x                   \\
                   & 50               & 1210
                     & 208.98                    & -5.79x                   \\
                   & 60               & 1210
                     & 213.78                    & -5.66x                   \\ \hline
\end{tabular}
}
\begin{tablenotes}
\footnotesize
   \item[1] The total data size of the weights transmitted during 500 rounds.
\end{tablenotes}
\end{threeparttable}
\end{table}

\textbf{Impact of model compression}: We conduct an ablation study to evaluate the effectiveness of our sparsification and adaptive weight quantization algorithm. We assess the impact of quantization over different satellite network sizes $C$ and cluster numbers $K$. The results in Table~\ref{tab:quantization} demonstrate that our model compression method significantly reduces the total volume of transmitted global models, while maintaining or even improving accuracy compared to other methods in some cases. For instance, when \(C=100\) and \(K=6\), the quantized global models achieve a compression ratio of approximately 7.34× compared to the original models during the entire uplink transmission. However, as the scale of the satellite network grows, this reduction ratio may decline. This is attributed to the fact that, each client receives a smaller portion of the data in larger networks, resulting in increased gradient fluctuations during local training. Consequently, a greater number of parameters must be quantized using 8 bits to maintain the accuracy of the global models. Additionally, for a fixed satellite network size and a certain number of communication rounds, there are variations in the quantization effect across different clusters. This is due to the effectiveness of the quantization process being influenced by the actual size of the satellite cluster and the convergence speed of the satellite clients.

%% file: 6_conclusion.tex
\section{Conclusion}
\label{sec:conclusion}

In this paper, we propose for the first time a semi-supervised FL framework for heterogeneous satellite networks to reduce processing time, improve energy efficiency, and handle unlabeled satellite data. We introduce a clustering-based PS selection algorithm to handle dynamic topology due to satellite mobility and unstable communication links. This algorithm optimizes network partitioning and strategically selects in-orbit PS nodes to reduce transmission delay. Additionally, an adaptive quantization method is introduced to reduce communication overhead and energy consumption within clusters. Experimental results show that, compared with FL-based baseline methods, our proposal significantly reduces processing time and energy consumption while ensuring model accuracy. Our future work aims to enhance data integrity and confidentiality in collaborative learning by integrating advanced privacy-preserving mechanisms such as differential privacy, further strengthening its applicability in security-critical applications.